\documentclass[10pt,journal ]{IEEEtran}

\usepackage[dvips]{epsfig}
\usepackage[dvips]{graphicx}
\usepackage[bookmarks=false]{hyperref}
\usepackage{cite}
\usepackage{amsmath,amssymb,amsfonts,latexsym,cases,enumerate}
\usepackage{srcltx,amsmath,cases}
\usepackage{color}
\usepackage{epsf,subfigure}
\usepackage{times,cite,psfrag}
\usepackage{multirow}
\usepackage{tabularx}
\usepackage{color}

\begin{document}

\title{\huge{Concatenated Raptor Codes in NAND Flash Memory }}

\author{Geunyeong Yu,~\IEEEmembership{Member,~IEEE,} and Jaekyun Moon,~\IEEEmembership{Fellow,~IEEE}
\thanks{Copyright (c) 2013 IEEE. Personal use of this material is permitted.
However, permission to use this material for any other purposes must be
obtained from the IEEE by sending a request to pubs-permissions@ieee.org.}
\thanks{Geunyeong Yu and Jaekyun Moon are with the Department of of Electrical Engineering, Korea Advanced Institute of Science and Technology (KAIST) Daejeon 305-701, Republic Korea (e-mail: tp10dal@stein.kaist.ac.kr, jmoon@kaist.edu). This work was supported by the National Research Foundation of Korea under grant no. 2013-13057165, the Ministry of Knowledge Economy of Korea under grant no. 10035202, and SK hynix. Geunyeong Yu has been partially supported by Samsung Electronics.}}

\markboth{To appear in Journal of Selected Areas in Communications in 2014}{Shell \MakeLowercase{\textit{et al.}}: Bare Demo of IEEEtran.cls for Journals}

\maketitle

\begin{abstract}

Two concatenated coding schemes based on fixed-rate Raptor codes
are proposed for error control in NAND flash memory.
One is geared for \textit{off-line recovery of uncorrectable pages}
and the other is designed
for \textit{page error correction} during the normal read mode.
Both proposed coding strategies assume hard-decision decoding of
the inner code with inner decoding failure generating erasure
symbols for the outer Raptor code.
Raptor codes allow low-complexity decoding of very long codewords while
providing capacity-approaching performance for erasure channels.
For the off-line page recovery scheme,
one whole NAND block forms a Raptor codeword with
each inner codeword typically made up of several Raptor symbols.
An efficient look-up-table strategy is devised for
Raptor encoding and decoding which avoids using large buffers in the controller
despite the substantial size of the Raptor code employed.
The potential performance benefit of the proposed scheme is evaluated
in terms of the probability of block recovery conditioned
on the presence of uncorrectable pages.
In the suggested page-error-correction strategy, on the other hand, a hard-decision-iterating
product code is used as the inner code.
The specific product code employed in this work is based on row-column
concatenation with multiple intersecting bits to allow the use of longer component codes.
In this setting the collection of bits captured within
each intersection of the row-column codes acts as the Raptor symbol(s), and the intersections of failed row codes and column codes are declared as erasures.
The error rate analysis indicates that the proposed concatenation provides a considerable
performance boost relative to the existing error correcting system based on long
Bose-Chauduri-Hocquenghem (BCH) codes.
\end{abstract}


%

\section{Introduction}

\IEEEPARstart{N}{AND} flash memories have become immensely popular due to lower power consumption, stronger shock resistance and higher data throughputs relative to hard-disk drives. Recently, high-capacity NAND flash memories
have been increasingly deployed in USB drives, solid state drives (SSDs) and all forms of mobile devices.
However, the high cost-per-bit remains the main weakness of NAND flash memories.
As the demand for cost reduction continues, multi-level cells (MLCs) which contain more than two bits per memory cell are also being widely employed. The MLC-based NAND flash can improve storage density, but the increased sensitivity
to noise presents a critical issue. Aggressive technology scaling has
also contributed significantly to the bit-cost reduction of NAND flash memory.
However, technology scaling also results in
the reduced noise margin and increased cell-to-cell interference.

To get around these issues, combinations of hardware and software techniques are typically employed.
The controller of NAND flash systems tries to distribute the program/erase cycles over the entire set of data blocks to prolong the lifetime of flash memories, a scheme called \textit{wear leveling} \cite{Wear-leveling}. A \textit{bad block management} technique is employed to keep track of bad blocks of which uncorrectable page errors have been encountered, preventing
writing of data to the known bad blocks.
Another essential element of
the modern storage system controller
is the error-correcting code (ECC)
used to ensure data reliability.

The trend in ECC is to employ increasingly strong codes with significantly improved error correction capability.
In fact, the raw channel bit error rate (BER) before error correction 
is now allowed to reach $10^{-3}$ and higher.
While long BCH codes are still very popular, the low-density parity-check (LDPC) \cite{Gallager:63MIT, Maeda:09DFT, Guiqiang:11CASI,Jiadong:11GLOBECOM} and turbo codes have emerged as promising candidates as well \cite{Berrou:ICC93}.
But both LDPC codes and turbo codes require soft read signals off the NAND chips to realize full performance potential,
and given the way all cells in a common word line is read simultaneously via comparison against
a uniform threshold level, generating soft read values for the NAND memory cells necessitates multiple reads with varying sensing levels \cite{Guiqiang:11CASI}.
Multiple reads directly translate to increased system latency and thus present difficulty in delay-sensitive applications.

While strong error protection is also possible using very long BCH codes,
decoding complexity increases rapidly with the error correction capability
and the codeword length. One way of utilizing the existing binary BCH codes
to strengthen page error protection is to apply product concatenation
of BCH codes along row and column directions of a two-dimensional buffer. In particular,
generalized product concatenation of BCH codes can allow
each pair of row and column codes to intersect at multiple bits
rather than at a single bit as in the traditional product code based on binary component codes \cite{Cho:12ICC}.
This block-wise product code (BW-PC) allows the use of stronger component codes
for a given overall codeword size while compromising the ability of locating
specific erroneous bit positions of the usual bit-wise product concatenation.

In this paper, we focus on concatenated codes based on Raptor codes, efficient erasure-correcting codes originally presented as a type of fountain codes ideal for
multi-casting or broad-casting in network communications. Fountain codes, including Raptor codes, used in the network communication setting are often described as ``rateless" in the sense that the number of encoder packets is not fixed before transmission. The encoder in these codes continues to generate coded packets (or symbols) as (pseudo) random linear combinations of source symbols and transmits them until all the intended receivers have recovered all the original source symbols. It is known that Raptor codes are asymptotically capacity-approaching in erasure channels whether the erasure probability are available at the transmitter or not \cite{Shokrollahi:06IT}. Practical and theoretic aspects of Raptor codes have been the subject of ongoing studies \cite{Byers:98ACMS,Byers:02JSAC,Venkiah:07ISIT,Saejoon:08CL,Mladeno:11Comp,Mahdaviani:12J_Com,Shokrollahi:11FTCI}.

We shall consider here the application of a specific Raptor code known as the R10 code \cite{Shokrollahi:11FTCI,Luby:07RFC,3GPP:MBMS} in a fixed-rate setting. This code exhibits optimal error rate performance as a random fountain code which has a decoding error probability of $2^{-(N'-K)}$ where $N'$ is the number of correctly received symbols and $K$ is the number of source symbols. The R10 code has two important desirable  characteristics: it is a systematic code and efficient decoding is possible based on what is called \textit{inactivation decoding}. Inactivation decoding employs belief propagation (BP) decoding as much as possible while also attempting to solve a linear system of equations when necessary, in an effort to realize efficient overall decoding
without compromising the performance of near-maximum-likelihood decoding. Because of its superior performance and linear-time encoding/decoding feature, the Raptor code has been adopted by the 3rd Generation Group Partnership Project
(3GPP) to be used in multimedia broadcast/multicast services (MBMS) for forward error correction \cite{3GPP:MBMS}
  and digital video broadcast-handheld (DVB-H) systems \cite{DVB-H}.
    The application of Raptor codes to binary-input memoryless symmetric channels in a fixed-rate setting has also been studied in \cite{Etesami:06IT,Cheng:ISIT07}.

In this paper, we specifically consider 1) off-line recovery of contaminated pages and 2) real-time page error correction, both based on concatenation of BCH-Raptor codes.
In current NAND flash memory systems, reading and writing are done on a page by page basis. Each page contains several kilo bytes while each NAND block is made up of 128 to 256 pages.
In our proposed off-line page recovery mode, each page is divided into several words.
A large number of such words from all pages within the given NAND block forms
a Raptor codeword with each word corresponding to one or more Raptor symbols.
In the proposed setting, several Raptor symbols also form an inner BCH codeword.
In this setting, the Raptor code attempt to clean up erasures that may arise as a result of the inner decoding failure. With typical code system parameters chosen, a significant number of  uncorrectable pages within a block
can be recovered with a very small extra coding overhead.

In some of current advanced NAND flash memory systems, extra protection
for the contaminated pages that cannot be
corrected by the deployed ECC (due to e.g. physical damages or wear in the block of data written overtime) is done via the redundant arrays of inexpensive disks (RAID) architecture based on the distribution of data and its parity across several NAND chips \cite{Balakrishnan:10ST, Soojun:11COM}. However, RAID systems require a high level of redundancy in
storing parity. In contrast, a Raptor code applied to the whole NAND block
provides protection against failed pages with a very small extra coding overhead while allowing for efficient encoding and decoding. The main challenge in our use of the Raptor code in the page recovery mode is that the codeword of the Raptor code corresponds to the entire NAND block,
which is in the order of mega bytes in size. While this size is way too large for a practical controller
to handle, our proposed scheme based on table-lookup strategies
allows encoding and decoding using only a small amount of buffers. Compared to the RAID system, our scheme does not allow real-time extra page
protection; the low-complexity and low-overhead features come at the cost of a slower processing speed.

On the other hand, in our proposed page-error-correction mode during normal page reads,
a Raptor code is concatenated with an inner BW-PC
based on relatively small BCH component codes.
Iterative hard-decision-decoding is utilized for the inner code.
In this setting the collection of bits captured within
each intersection of the row-column codes acts as one or more Raptor symbols, and the intersections of failed row codes and  failed column codes after BW-PC decoding are declared as erasures
for the outer Raptor code. Here, each page basically corresponds to a single Raptor codeword.
Our error rate analysis indicates that the proposed concatenation provides a considerable performance boost relative to the existing BCH-based coding schemes.

Note that while the Reed-Solomon (RS) codes are also a clear option as the outer code,
decoding complexity grows quickly with the increasing block size. 
The RS codes are deemed a less attractive solution for the problems at hand. 

The term ``block" in the block-wise product code (BW-PC) of \cite{Cho:12ICC}
is not to be confused with the term ``block" in the typical description of the  NAND block structure.
To avoid ambiguity,
the block of bits captured in the intersections of row and column codewords
will also be referred to as symbol in this paper, although the context will usually make
the intended meanings clear.

The paper is organized as follows. In Section \ref{S:Raptor_review} we briefly review Raptor codes. In Section \ref{S:Raptor_block_recovery} we describe the Raptor-code-based off-line page recovery scheme.
The detailed encoding and decoding algorithms of the Raptor code used in the page recovery mode are described and the error rate performance is evaluated.
In Section \ref{S:RC_BW_TPC} the second concatenated Raptor code geared for real-time page error correction is discussed. It is
shown that the channel is effectively converted from a binary symmetric channel (BSC) to a symbol-erasure channel via the use of the inner BW-PC.
The error rate performance for the BW-PC and Raptor decoding is also evaluated in this section.
Finally, we conclude our work in Section \ref{S:Conclusion}.

\section{Background}\label{S:Raptor_review}

\subsection{Fountain Codes}

Fountain codes are a type of erasure codes first proposed for multi-cast network communications \cite{Byers:98ACMS,Byers:02JSAC}.
Given a source data consisting of $K$ symbols (frequently referred to as packets in the fountain code literature), the encoder of a fountain code
produces a stream of output symbols, each of which is generated independently
as some linear combination of the source symbols with the linear mapping chosen anew each time in a pseudo-random fashion. It is possible to recover the original $K$ symbols from any set of $N=K(1+\epsilon)$ received symbols with high probability (at least $1-1/K^{c}$), where $c$ is a constant. The number of encoded packets is not fixed beforehand
and transmission typically continues until all receivers (in the multi-cast mode) recover all source symbols.
An encoder potentially generating an endless stream of packets in all directions
until all receivers are satisfied conjures up the image of a fountain, hence the name fountain codes.
Also, this class of codes is often referred to as rateless since the number of encoded packets is not fixed before the transmission, which in the context of traditional error control coding
is to say that the code rate is not predetermined.
At the transmitter and the receiver resides the same generator matrix for the linear mappings of the source symbols
to the transmitted packets. When the coded packets or
symbols are transmitted, the receiver throws out the erased symbols and
reduces the generator matrix accordingly before decoding the original source symbols.

From the complexity view point, to encode each symbol requires $K/2$ symbol additions; since
a total of $N\approx K$ coded symbols are transmitted, encoding complexity can be described as $K(K/2)$.
A conceptually simple decoding strategy would be to perform a matrix inversion and, using
Gaussian elimination, this will cost about $K^3$ symbol additions.

Luby transform (LT) codes represent the first practical
realization of fountain codes \cite{Luby:02FCS}. LT codes use the low-density generator matrix (LDGM) and message-passing decoding to avoid direct matrix inversion. With a careful choice of the node degree distributions in the code's factor graph, it has been shown that the LT decoder can recover all $K$ source symbols with small coding redundancy.
However, a significant drawback remains: namely, the average node degree is $O(K\ln{K})$, which does not quite allow
linear-time encoding/decoding.

Raptor codes \cite{Shokrollahi:06IT}, an extension of LT codes, are designed to achieve linear-time encoding/decoding. Raptor codes employ an outer code concatenated with a weakened LT code which has a significantly
low average node degree. An LT code with a lower average node degree is normally problematic
since there exists a significant probability that a source symbol may not be connected to clean received symbols
in the factor graph, a situation that leads to decoding failure. But in Raptor codes,
these unconnected source symbols can be corrected by an outer code with a very small coding redundancy.
Overall, Raptor codes can achieve nearly linear-time encoding/decoding and has been proved to be
asymptotically (as $K$ tends to infinity) capacity-achieving on a binary erasure channel (BEC) without any channel information at the transmitter or at the receiver \cite{Shokrollahi:06IT}.

\subsection{Raptor Encoding and Decoding}\label{Ss:Raptor enc./dec.}

A fountain code can also be viewed as a regular linear block code which can be represented by a generator matrix. This view facilitates exposition of the encoding and decoding processes of Raptor codes.
We specifically describe encoding and decoding of the R10 code, which is a systematic code and allows
simple evaluation of the error rate performance for low erasure-rate channels.
The description in this subsection is basically a quick overview of the algorithms given in  \cite{Luby:07RFC, Shokrollahi:11FTCI}.

\vspace{4 mm}
\noindent \textbf{Encoding process}

There are basically two steps in the encoding process. The first step is to generate some $L$-intermediate symbols from the source symbols. Let us assume that the code length is $N$. Let $\mathbf{m}$ denote the vector of $L$ intermediate symbols and let $\mathbf{t}=[\mathbf{z}^{T}\  \mathbf{s}^{T} ]^T$ where $\mathbf{z}$ is a vector of $(S+H)$ zero symbols and $\mathbf{s}$
is a vector of $K$ source symbols with $(\ )^T$ indicating transpose.
Then $\mathbf{m}$ and $\mathbf{t}$ are related by $\mathbf{A}_{pre}\mathbf{m}=\mathbf{t}$,
 where $\mathbf{A}_{pre}$ is an $L\times L$ matrix over $GF$(2) and $L=K+S+H$.
  Here, the first $(S+H)$ rows of $\mathbf{A}_{pre}$ describe the relationship among the intermediate symbols and the last $K$ rows contain connection information between the intermediate symbols and source symbols, consistent with
   the designed degree distribution. Hence, the intermediate vector $\mathbf{m}$ can be produced as:

   \begin{equation} \label{EQ:produce C}
   \mathbf{m} = \mathbf{A}_{pre}^{-1} \mathbf{t}.
   \end{equation}

Once the intermediate symbols are determined, then in the second step $N-K$ redundant symbols are encoded by the LT relationships depicted by $\mathbf{G}_{LT}$, which is an $M\times L$ binary matrix with $M=N-K$. The LT generator matrix can produce any number of parity symbols $\mathbf{r}$ according to
\begin{equation}
    \mathbf{r} = \mathbf{G}_{LT} \mathbf{m},
\end{equation}
and the coded vector $\mathbf{c}$ can be written as
\begin{equation}\label{Eq:Re-generation}
    \mathbf{c} = \mathbf{A} \mathbf{m}= [\mathbf{z}^T\  \mathbf{s}^T \ \mathbf{r}^T]^T\\
\end{equation}
where
\begin{equation}
    \  \mathbf{A}=
    	\begin{bmatrix}
    	\mathbf{A}_{pre} \\
    	\mathbf{G}_{LT} \\
    	\end{bmatrix}.
\end{equation}
Here, we note that $\mathbf{A}$ is known to both the encoder and the decoder and the value of $M$ is selected to be sufficiently large to compensate for possible loss of encoded symbols in the channel.

\vspace{4 mm}
 \noindent \textbf{Decoding process}
 
After the coded vector (excluding the first $S+H$ zeros symbols) is transmitted through the channel, decoding also consists of two basic steps - the first step for finding the intermediate symbols and the second step for recovering the erased symbols using the original generator matrix $\mathbf{A}$.
Let us assume that $\mathbf{c}'$ is the vector of successfully received symbols followed by a length-($S+H$) all-zero vector, $N'$ is its length and $\mathbf{A}'$ is  the reduced generator matrix obtained from $\mathbf{A}$ by eliminating the rows corresponding to the erased symbols.
Therefore, the success of decoding depends on whether $\mathbf{A}'\mathbf{m}=\mathbf{c}'$ can be solved or not.
If the intermediate symbol vector $\mathbf{m}$ can be obtained via matrix inversion,
then in the second step the erased symbols of $\mathbf{c}$ can be reproduced using $\mathbf{A}$ and $\mathbf{m}$.

\vspace{4 mm}
 \noindent \textbf{Matrix Inversion and Inactivation Decoding}

 Decoding and encoding of Raptor code amount to solving the equations
  $\mathbf{A}'\mathbf{m}=\mathbf{c}'$ and $\mathbf{A}\mathbf{m}=\mathbf{t}$, respectively, as described above.
  To obtain $\mathbf{m}$ from $\mathbf{c}'$, Gaussian elimination (GE) can be performed.
  Efficient decoding algorithms have been suggested for this
  in \cite{3GPP:MBMS}, \cite{Shokrollahi:11FTCI} and
  an improved implementation of maximum-likelihood (ML) decoding algorithm has also been presented in \cite{Saejoon:08CL} to solve the same problem.
  These algorithms are different mainly in their ordering of the rows for Gaussian elimination, but they all give the same result. The objective is to turn $\mathbf{A}'$ into an identity matrix by row exchanges, column exchanges and row additions.

The decoding algorithm can also be described by the \textit{belief propagation} (BP) decoding \cite{Gallager:63MIT,Luby:01IT,Luby:97ATC }. In the case of the erasure channel, the BP algorithm is best described in terms of the ``decoding graph'' corresponding to the relationship between the received coded symbols and the intermediate symbols. This is a bipartite graph between the $L$ intermediate symbols and the $N'$ received encoded symbols. Note that the BP algorithm for general LT or Raptor codes can be described \cite{Mackay:05IEEC} as: 1) finding a degree-1 encoded symbol, 2) delivering the value of the chosen encoded symbol to the connected source symbol (or intermediate symbol in the case of a Raptor code), 3) exclusive-ORing the given values to all the encoded symbols adjacent to the selected source (or intermediate) symbol and 4) eliminating all edges emanating from the source (or intermediate) symbol. The steps 1) though 4) are performed iteratively until there remain no unrecovered source (or intermediate) symbols
or there is no degree-1 encoded symbol left in the graph. In comparison,
in the decoding algorithm geared to the R10 code, the decoding process is effectively allowed to continue
even if there is no degree-1 encoded symbol in the graph
via a special procedure called \textit{inactivation decoding} \cite{Shokrollahi:11FTCI}.
This can be envisioned as allowing the BP algorithm to continue by removing (or inactivating) certain edges and nodes even after
the BP algorithm gets stuck in the conventional graph.
Once BP decoding halts, a simple linear system of equations arises 
according to the inactivated notes and edges; if there exists a solution, then the whole intermediate symbols can be recovered. It can be shown that decoding is possible overall whenever the reduced matrix $\mathbf{A}'$ is full rank. Assuming a matrix size of $q$, solving it requires exclusive-OR operations in the order of $q^2$. 
Thus, if $q$ is in the order of $\sqrt{K}$, then the computational cost can be said to be linear in $K$.
The R10 code of \cite{Shokrollahi:11FTCI} is actually designed to satisfy this condition,
requiring $O(K)$ symbol additions for overall decoding.

\section{Raptor Code in Block Recovery}\label{S:Raptor_block_recovery}
In this section, we introduce a Raptor-code-based concatenation scheme where each codeword of the inner code encompasses several Raptor symbols. In this setting the entire NAND block of flash memory cells become a single Raptor codeword.
 The channel model of NAND flash is assumed to be a BSC in this paper since all decoding is based on hard-decision inputs. The overall code construction for the proposed scheme along with detailed
  encoding and decoding algorithms are described, and the error rate performance is evaluated.

The simplest way to convert a BSC to a packet-erasure channel is to use an existing $(n,k)$ inner error detection or correction code, where $n$ is the codeword length and $k$ is the information word length.
Any linear block codes such as the LDPC, BCH code and Reed-Solomon (RS) codes can be used
as the inner code. While the error detection code in general allows a high overall code rate, the erasure rate
of the converted channel can be excessively high. In our investigation, inner ECCs with a reasonably high rate
have proven to be a better choice. Our discussion in this paper will focus on the use of inner ECCs.

\begin{figure}[t]
\begin{center}
\centering
	\includegraphics[width=0.4\textwidth]{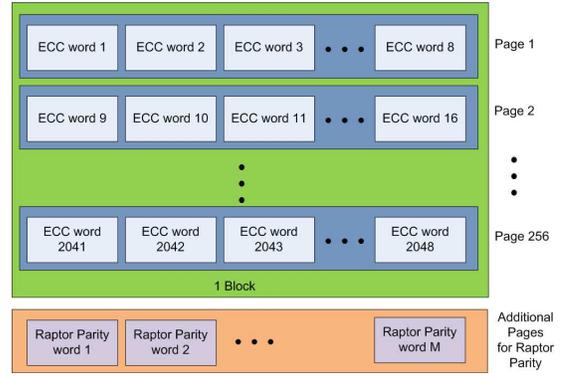}
\caption{Raptor Codes in NAND flash memory for the block recovery ($N_{s}$=1, $p$=256, $w$=8)}
\label{FIG:Raptor_System}
\end{center}
\end{figure}

We will specifically use the BCH codes as inner codes. We assume that one NAND block consists of $p$ pages and one page contains $w$ inner codewords.
We also assume that each codeword of the inner BCH code exactly contains $N_{s}$ source symbols of the Raptor code.
Fig. \ref{FIG:Raptor_System} shows an example of the Raptor code applied to page  recovery when $N_{s}=1$, $p=256$ and $w=8$. Let $K$ be the number of inner ECC codewords  in one block (i.e., $K=w p$), and the total size of the Raptor parity symbols
be equal to that of $N-K$ inner codewords.
The Raptor codeword contains $nN$ coded bits corresponding to $kK$ information bits, and there exist $N_{s}N$ Raptor symbols consisting of $N_{s}K$ source symbols and $N_{s}(N-K)$ parity symbols. In the overall decoding process, $N_{s}$ symbols are declared as erasures whenever the inner decoder fails, and the Raptor decoder then attempts to recover the erased source symbols.

\subsection{Encoding and Decoding Algorithm}\label{Ss:enc_dec}

As mentioned earlier, the NAND block size is in the order of mega bytes, and direct encoding and decoding
of a Raptor code of this size are not feasible due to the required memory buffer size.
To this end, we devise an efficient algorithm which uses only small buffers for the encoding and decoding processes.
The key idea is that the lookup table of each parity symbol can be pre-calculated in the form of the position vector which contains the addresses of the required source symbols to generate the given parity symbol. Likewise in the proposed decoding process, the look-up tables for the erased symbols declared after inner decoding also are pre-calculated in the forms of the position vectors of the required received symbols. 

In the encoding process, as the input data stream gets to the NAND block\footnote{Notice that although our exposition encourages visualization of coding on the entire physical NAND block, our scheme actually employs logical blocks so that codewords may not necessarily come from a common physical NAND block.}  
the required source symbols are exclusive-ORed to calculate each parity symbol according to the pre-calculated look-up table.
In decoding, if the recovery mode is invoked by the corrupted words or pages, 
only the required symbols are read to recover the erased symbols. 
This is possible because the parity symbols and the erased symbols are simple modulo sums of known sets of the source symbols and the encoded symbols, respectively.
We note that the look-up tables of parity symbols are pre-calculated only once and used repeatedly, but the look-up tables of erased symbols need be calculated according to the  positions of the erasures.

The look-up table of the erased symbols can be obtained in the proposed decoding process as follows. Decoding of a Raptor code basically consists of solving
\begin{equation}
     \mathbf{A}'\mathbf{m}=\mathbf{c}'
     \label{R_decoding}
\end{equation}
for $\mathbf{m}$ and then regenerating the erased symbol using $\mathbf{m}$ and $\mathbf{A}$ as described in the subsection \ref{Ss:Raptor enc./dec.}.
To solve (\ref{R_decoding}), Gaussian elimination should be performed which turns $\mathbf{A}'$ into the form of an identity matrix by row exchanges, column exchanges and row additions. Whenever two rows of $\mathbf{A}'$ are exchanged, the corresponding symbols of $\mathbf{c}'$ are also exchanged, and when two columns of $\mathbf{A}'$ are exchanged, the corresponding symbols of $\mathbf{m}$ are also exchanged. When the $i$-th row of $\mathbf{A}'$ is added to the $j$-th row, the $i$-th symbol of $\mathbf{c}'$ is also added to the $j$-th symbol. This process can be traced 
by multiplying the identity matrix to right side of (\ref{R_decoding}):
\begin{equation}
     \mathbf{A}'\mathbf{m}=\mathbf{I}_{N'}\mathbf{c}'
     \label{Look-up}
\end{equation}
and then transforming (\ref{Look-up}), through Gaussian elimination, to:
\begin{equation}
     \mathbf{I}_{L}\mathbf{m'}=\mathbf{P}\mathbf{c}'
     \label{Look-up2}
\end{equation}
where
$\mathbf{m'}$ is the reordered intermediate symbol vector. Then, $m'_i$, the $i$-th 
symbol of $\mathbf{m'}$ can be written as $\mathbf{p_ic'}$, where $\mathbf{p_i}$ is a row vector corresponding to the $i$-th row of $\mathbf{P}$.
The erasure symbols can finally be reconstructed from  $\mathbf{c} = \mathbf{A} \mathbf{m}$.

Let the look-up table for the $k$-th parity symbol be $\mathbf{b}_{k}^p$ with the $l$-th component
  $b_{k,l}^p$ set to `1' when the $l$-th source symbol is necessary to generate the given parity symbol, and to `0' otherwise.  
  Likewise, let the look-up table for the erased symbol corresponding to the $j$-th coded symbol be the binary vector $\mathbf{b}_{j}^e$ with the $i$-th component $b_{j,i}^e$ set to `1' when the $i$-th received symbol is necessary to recover the erased symbol, and to `0' otherwise.

Let us consider a specific example where the equality $\mathbf{A} \mathbf{m}=\mathbf{c}$
is given by: 

  \begin{equation}\label{EQ:ex1}
   \left[ \begin{array}{c c c c } 0& 1&0&1\\ 1&0&1&1 \\ 0&1&0&0 \\ 0&0&1&1 \\1&0&0&1 \end{array} \right]\begin{bmatrix} m_1\\m_2\\m_3\\m_4\end{bmatrix} = \left[ \begin{array}{c} s_1 \\s_2\\s_3\\s_4\\r_1 \end{array} \right]
   \end{equation}
where the last source symbol $r_1$ represents the sole parity symbol.  
Assume that $s_4$ is erased in the channel. Then, after eliminating $s_4$, the equation 
 $\mathbf{A}'\mathbf{m}=\mathbf{c}'$ becomes
 (\ref{EQ:ex1}) reduces to
  \begin{equation}\label{EQ:ex2}
   \left[ \begin{array}{c c c c } 0& 1&0&1\\ 1&0&1&1 \\ 0&1&0&0 \\1&0&0&1 \end{array} \right]\begin{bmatrix} m_1\\m_2\\m_3\\m_4\end{bmatrix} = \left[ \begin{array}{c} s_1 \\s_2\\s_3\\r_1 \end{array} \right].
   \end{equation}

To obtain the look-up table for the erased source symbol $s_4$, Gaussian elimination is executed which turns $\mathbf{A'}$ in (\ref{EQ:ex2}) into an identity matrix by row/column exchanges as well as row additions. The transformed equality $\mathbf{I}_{L}\mathbf{m'}=\mathbf{P}\mathbf{c}'$ is now

    \begin{equation}\label{eq:1}
   \left[ \begin{array}{c c c c } 1& 0&0&0\\ 0&1&0&0 \\ 0&0&1&0 \\0&0&0&1 \end{array} \right]
   \begin{bmatrix} m_2\\m_4\\m_1\\m_3\end{bmatrix} =
      \left[ \begin{array}{c c c c } 0& 0&1&0\\ 1&0&1&0 \\ 1&0&1&1 \\1&1&1&1 \end{array} \right]
     \left[ \begin{array}{c} s_1 \\s_2\\s_3\\r_1 \end{array} \right].
   \end{equation}
   
The intermediate symbols can be expressed as $m_2=\mathbf{p_1c'}$, $m_4=\mathbf{p_2c'}$ and so on.
 To recover $s_4$, we go back to (\ref{EQ:ex1}) to find that $s_4$ is sum of $m_3$ and $m_4$.
 Hence, $s_4$ is now reconstructed as $\mathbf{(p_2+p_4)c'}$ and the corresponding look-up table is $[0 1 0 0 1]$, indicating that the second and fifth encoded symbols are used to recover the given symbol.

Note that the off-line recovery mode is invoked only in the presence of uncorrectable inner codewords. The overall system diagram of the Raptor encoder and decoder is shown in Fig. \ref{FIG:Raptor_Framework}. 
For the case where each word protected by the inner code within a block is a single Raptor symbol (i.e. $N_s=1$), the overall encoding and decoding steps are summarized as follows:

\begin{figure}[t]
\begin{center}
\centering
	\includegraphics[width=0.4\textwidth]{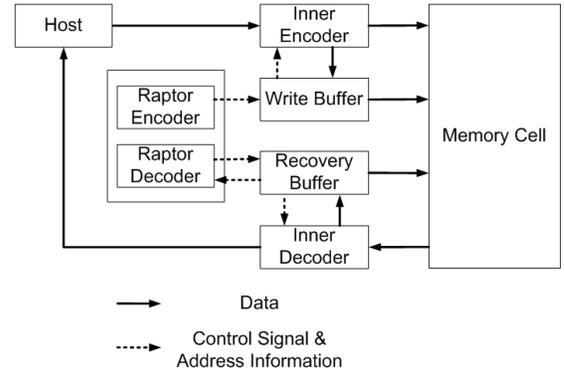}
\caption{Overall System Diagram for the Raptor encoder and decoder.}
\label{FIG:Raptor_Framework}
\end{center}
\end{figure}

\vspace{4mm}
\noindent \textbf{Encoding process}

\begin{enumerate}[ 1.]
\item Before the data is written into the NAND flash memory cells, the Raptor encoder generates the look-up tables $\mathbf{b}_k^p$ for $k= 1,2,...,N-K$) and then stores them, in the write buffer.

\item While the data is being written into the memory cells, for the $l$-th incoming word $W_{l}$,

	a) the write buffer checks if $b_{k,l}^p$ is `1' for all $k$.
	
	b) if true, the parity symbol $P_k$ (initially set to all zeros) is updated after being exclusive-ORed with $W_{l}$.
	
	c) repeat steps a) and b) for all $l$ values.

\item When the $j$-th word of a given NAND block is updated, the write buffer updates the appropriate parity words
by exclusive-ORing them with the new word.

\vspace{0.8mm}

\end{enumerate}

\noindent \textbf{Decoding process}
\begin{enumerate}[ 1.]
\item While reading the data from the memory cells, if uncorrectable errors occur which cannot be recovered by the inner codes,  the recovery buffer marks them as the erasure words, and saves their positions. The position information for the erasure words are delivered to the Raptor decoder.

\item For each erasure word $E_j$ in position $j$, the look-up table $\mathbf{b}_j^e$ is generated and delivered to the recovery buffer.
\item When the off-line recovery mode is invoked, necessary 
words are read from the memory cells and cleaned up using the inner decoder.
For the resulting $i$-th word $W_i$,

a) the recovery buffer checks if $b_{j,i}^e$ is `1' for all $j$.

b) if true, $E_j$ is exclusive-ORed with $W_i$.

\item After all erased symbols $E_j$ are recovered, they are written to new logical memory cell locations. 

\end{enumerate}
\vspace{4mm}

The proposed recovery scheme based on Raptor codes can recover the corrupted words which cannot be recovered by the inner codes. By utilizing the  pre-calculated look-up tables, efficient encoding and decoding are possible without directly handling a block of data, the size of which is in the order of mega bytes, making direct encoding and decoding practically infeasible. Here, we should point out that the proposed scheme requires re-writable memory for maintaining and updating parity symbols as well as extra memory to store the look-up tables for encoding and decoding. But in our proposed system setting, the parity portion in a block represents a very small percentage of the block and the size of the look-up tables for Raptor decoding is $N\times N_E$ bits, where $N_E$ is the number of erased symbols, which should be negligible compared to the size of one block.

We have also examined required computation complexity of Raptor decoding in the proposed recovery mode. It is known that decoding complexity of $K$ Raptor symbols requires $O(K)$ \textit{symbol} additions \cite{Shokrollahi:06IT}. In the proposed scheme, there is an additional computation burden to construct the look-up tables, requiring a total of $O(K\sqrt{K})$ \textit{binary} additions. Since the size of the Raptor symbol in the proposed recovery scheme is in the order of 1KB, when the number of source symbols is set to 8000, the additional amount of computation can be roughly estimated as hundreds of symbol additions. This represents an increase in the overall computation by less than 2\% of that required for original Raptor decoding.

\vspace{2mm}
\noindent \textbf{Practical Data-Management Issues}

Recall that the proposed scheme is applied to logical blocks. When one logical block is written,
data in this block is typically distributed to multiple physical blocks. In this scenario, some problems arise. These problems and their possible solutions are described as follows.

1) While data is written, parity symbols should be updated repeatedly. Thus, before the writing of one block of data is finished,
the corresponding parity symbols are maintained in a rewritable memory, which we assume to be static random access memory (SRAM).
After one block of data is completely written, the updated parity symbols are then written in the flash memory cells.
Note that SRAM is not needed for all physical blocks; the number of parity symbol sets that need be maintained in SRAM
is the same as the maximum number of parallel input streams handled by NAND flash.

2) When data writing is suspended for the given block with only a portion of NAND block filled up,
the calculated parity symbols can be written to NAND flash memory right away,
or they are held in SRAM until the block gets filled up later on.
The former method can reduce the use of SRAM, but when the rest of the data are written in the block,
the written parity symbols should be updated, which would mean invalidating the pages that contain
the existing parity symbols. This would cause more wasted flash memory space as
new parity symbols must be written onto new physical pages. The latter approach, on the other hand,
can avoid the waste of space, but SRAM is required to house the parity symbol sets for all unfilled blocks, which would not be practically feasible.
In reality, a judicious mix of two strategies would be desirable.

3) When some pages of the block are updated, new data must be written to new physical pages and the out of date pages are declared invalid.
In this case, to update the parity symbols, the invalid pages and the current parity symbols
 should be read. In the use of RAIDs, this is also an inevitable problem.
 To reduce the cost of parity update, methods such as Partial Parity Cache (PPC) \cite{Soojun:11COM} and Flash-aware Redundancy Array (FRA) \cite{Lee:09CI} have been proposed, which can also be employed for our purposes here.

 In SSDs, the concept of \textit{internal parallelism} is built-in to increase the bandwidth of reading and writing. In an SSD, flash memory chip packages are connected to the controller through multiple channels, and each flash memory package includes two or more chips. Each chip can be selected individually and perform a read, write or erase operation independently. Hence, when there exist 16 channels and 4 chips per package, for example, the internal parallelism degree will be 64 and the maximum number of the partially filled blocks will be 64. Thus, in the proposed scheme, to solve the problems described in scenarios 1) and 2), the whole size of the Raptor parity symbol sets that need be maintained in SRAM is 64 times the size of the Raptor parity symbol. This would mean that the required SRAM buffer size is less than 2.6 MB, assuming that the block size is 2MB and the Raptor code has a 2\% coding overhead.

Before ending this subsection, we remark that because the proposed block recovery method actually deals with logical blocks rather than physical blocks, the idea may be utilized for recovery of data stored across multiple NAND chips, somewhat akin to RAID. The potential issue that needs be carefully addressed here, however, is that the size of the blocks needs be large enough to ensure the efficiency of the Raptor code, which may lead to system-level challenges for handling long decoding latency.

\subsection{Performance Evaluation} \label{Ss:system_perform_recovery}

Recall that one inner codeword covers $N_s$ Raptor symbols exactly, $K$ is the number of inner codewords in a block, and the bit size of $N-K$ inner codewords are the same as that for the $N_{s}(N-K)$ parity symbols. The failure rate of the Raptor decoder, given that the number of the failed inner codes $N_E$ is $i$ and $N_s\cdot i$ erasure symbols are produced, is given by

\begin{equation}
P(E_1 | N_E = i) = \min\{1,2^{-(N_{s}(N-K)-N_{s}\cdot i)}\}
\end{equation}
making use of the expression developed in \cite{Shokrollahi:11FTCI}. The R10 code
has performance close to that of a random binary fountain code.
So, when the receiver collects $C$ ($=S+R$) symbols except the erased symbols, where $S$ is the number of source symbols to be decoded, the probability of \textit{decoding failure} of the R10 code is roughly $2^{-R}$, which is also the probability that a $C$ by $S$ random binary matrix is full rank. The $\min$ function is used because when $R$ is less than 0 the decoding failure rate is 1.

 When the $(n,k,t)$ BCH code is used, the erasure probability of the inner code can be obtained by

 \begin{equation}\label{EQ:Erasure}
P_E =1-\sum_{i\leq t}{{n\choose i}{p_e}^i (1-p_e)^{n-i}}
 \end{equation}
where $p_e$ is the raw bit error rate (BER).
  Then the error probability $P_F$ of the proposed concatenated codes is the probability of the events that decoding fails in the Raptor code (event $E_1$) or any miss correction occurs over the whole words in the block (event $E_2$). It can be easily shown that $P_{F}$ satisfies:

\begin{equation}\label{EQ:P_BLER_Anal_Reduced}
\begin{aligned}
&P_{F} \\
&= P\{E_1 \ \ or \ \ E_2\} \approx P\{E_1\} + P\{E_2\} \\
&=  \sum_{i=1}^{i=K}P(E_1 | N_E=i)P(N_E=i) +1-\left(1-P_{u.e.}\right)^K  \\
&\leq \sum_{i=1}^{i=K}{\left[{K\choose i}{P_{E}}^i (1-P_{E})^{K-i} \min\{1,2^{-(N_{s}(N-K)-N_{s}\cdot i)}\}\right]} \\
&{  }+K\cdot P_{ue},
\end{aligned}
\end{equation}
where $P_{ue}$ is the undetected error (miss correction) probability of the inner ECC. Hence, if the miss-correction probability and the erasure probability of the ECC are known, then the error rate performance can be predicted.

Equation (\ref{EQ:P_BLER_Anal_Reduced}) clearly confirms that to achieve a low decoding failure rate for the Raptor code, the miss correction probability of the inner code should be low and the number of parity symbols of Raptor codes should be sufficiently large. To satisfy both conditions with a high code rate, a large enough data size is required. One NAND block contains data in the order of mega bytes, satisfying 
both conditions when the Raptor code is applied for block recovery.

\begin{figure}[t]
\begin{center}
\centering
	\includegraphics[width=0.5\textwidth]{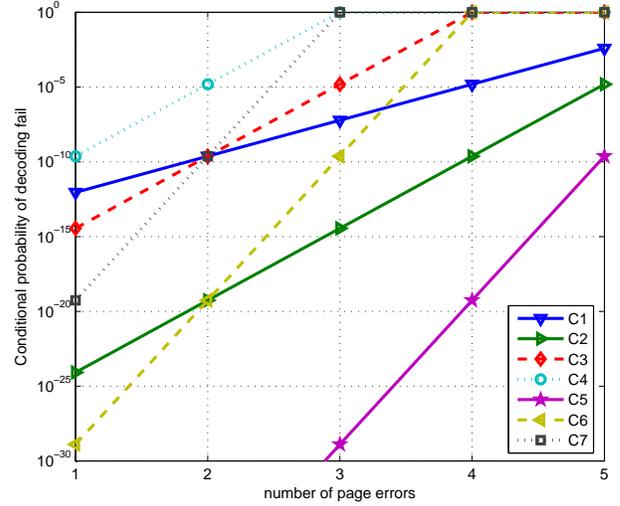}	
\caption{Conditional probability of Raptor decoding failure given the number of bad pages in a block for the 7 codes specified in Table \ref{TAB:Overhead_R10}. Solid lines are for the codes with a 2.34\% overhead ($C_1 , C_2$, and $C_5$) while dashed lines correspond to the codes with a 1.56\% overhead ($C_3, C_6$), and dotted lines are for the codes with a 1.17\% overhead ($C_4, C_7$).}
\label{FIG:Raptor_performance}
\end{center}
\end{figure}

\begin{table}[b]

\caption{}{Different Raptor code parameters with different symbol sizes and numbers of parity symbols. $N_{s}$ represents the number of Raptor symbols within each inner codeword of length 1 KB. SS and PS denote source symbols and parity symbols, respectively. The overhead of 2.34\% means additional 6 pages are used as the parity pages; the overhead of 1.56\% and 1.17\% mean 4 and
3 parity pages, respectively.} \vspace{0.1cm}
\begin{center}
\begin{tabular}{c||c|c|c|c|c} \hline
Cases& $N_{s}$  & No. of SSs    & Symbol size & No. of PSs & Overhead \\ \hline
$C_1$&1  & 2048               &  1   KB   & 48   & 2.34\%  \\ \hline
$C_2$&2  & 4096               &  1/2 KB   & 96   & 2.34\%  \\ \hline
$C_3$&2 & 4096               &  1/2 KB   & 64   & 1.56\%  \\ \hline
$C_4$&2 & 4096               &  1/2 KB   & 48   & 1.17\%  \\ \hline
$C_5$&4 & 8192               &  1/4 KB   & 192   & 2.34\%  \\ \hline
$C_6$&4 & 8192               &  1/4 KB   & 128   & 1.56\%  \\ \hline
$C_7$&4 & 8192               &  1/4 KB   & 96   & 1.17\%  \\ \hline

\end{tabular}\vspace{-0.1cm}
\label{TAB:Overhead_R10}
\end{center}
\end{table}

For the system performance, we have checked the probability of decoding failure in Raptor decoding conditioned on a variable number of uncorrectable pages.
We assume that one block consists of 256 pages with each page having 8 KB of user data corresponding to 8 inner codewords of size 1KB each.
In this setting, varying $N_S$ from 1 to 4, and the Raptor coding overhead from 1.17\% to 2.34\%, we have designed seven Raptor codes ($C_1 - C_7)$ with the parameters given in Table \ref{TAB:Overhead_R10}. With the extra coding overhead defined as the ratio of the number of Raptor parity pages to the number of source pages, the overheads of 2.34\%, 1.56\% and 1.17\% mean additional 6, 4 and 3 pages for Raptor parity, respectively.

For the chosen code parameter selections, the analytical results for the Raptor decoding failure rate are summarized in Fig. \ref{FIG:Raptor_performance}. For $C_1 , C_2$ and $C_5$, which utilize a 2.34\% overhead, the performance curves are shown in solid lines, and for the codes utilizing a 1.56\% overhead ($C_3, C_6$) performances are represented by dashed lines. Given the same overhead, as $N_s$ is increased, the decoding performance improves since the Raptor codes can utilize more parity symbols. For $C_5$, which use $N_s=4$ and a 2.34\% overhead, the Raptor decoder can recover the block for up to 5 failed pages with conditional
decoding failure rates of better than $10^{-10}$. For codes with the same $N_s$ but with
smaller coding overheads, $C_6$, with a 1.56\% overhead, can achieve conditional decoding failure rates below $10^{-10}$ up to 3 corrupted pages and $C_7$, with a 1.17\% overhead,
can correct 2 pages with a conditional failure rate of $10^{-10}$.

\section{Raptor Codes with {Block-Wise} Product Codes}\label{S:RC_BW_TPC}
In this section, the second Raptor-based concatenation scheme designed for page error correction during the normal read mode is
discussed. The proposed scheme uses the  BW-PC of \cite{Cho:12ICC}
as the inner code to declare erasure for the shared bits among failed row codes and
column codes after inner decoding. The overall code structure and the code parameter selection process are described.
Performance analysis along with simulation results are provided.

\subsection{Code Structure}\label{Ss:System_B_TPC}
The BW-PC consists of row and column
constituent codes that share multiple coded bits in their intersections
\cite{Cho:12ICC}. In this sense, this code can be viewed as a generalization of the conventional row-column product code
where every pair of row and column codes shares one bit. The advantage of the block-sharing is that for a fixed-size two-dimensional array,
larger (and thus stronger) row and column codes can be employed. The disadvantage is that when a given pair of row-column codes fail to decode,
error locations can only be identified down to the shared block of the failed row and column.
The work of \cite{Cho:12ICC} has shown that with iterative hard-decision decoding, an error floor appears
in the mid-to-low BER region due to dominant error events. In \cite{Cho:12ICC}, this error floor has been lowered significantly by iterative reliability-based decoding, with the reliability information acquired based on memory-sensing via multiple reads.
In this work, we instead apply an outer Raptor code to lower the error floor without resorting to multiple reads.

\begin{figure}[t]
\begin{center}
\centering
	\includegraphics[width=0.4\textwidth]{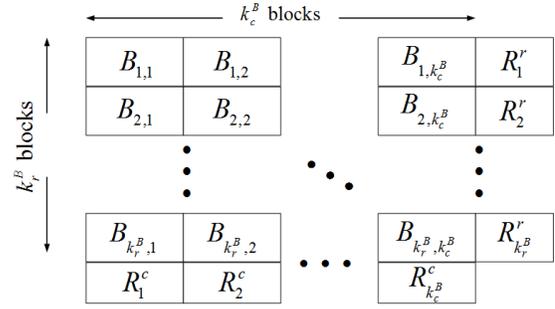}

\caption{Structure of BW-PC}
\label{fig:BW_TPC}
\end{center}
\end{figure}

For the BW-PC, binary BCH codes are considered as component codes. Let the codeword length, message length and error-correcting capability of
the row and column codes be characterized by $(n_{r},k_{r},t_{r})$ and $(n_{c},k_{c},t_{c})$, respectively.
Let $B_{i,j}$ denote a block of shared bits between the $i$-th row code and $j$-th column code. Let
$n_{B}$ be its size in bits, which is a key design parameter. 
This block of $n_B$ bits act as one or more ($N_i$) Raptor symbols in our setting.
 See Fig. \ref{fig:BW_TPC}. The symbols $R_{i}^{r}$ and $R_{j}^{c}$ represent the
blocks of parity bits for the corresponding row and column binary BCH codes, respectively. Let $m_r$ and $m_c$ be the respective parity lengths of the row and the column codes such that $m_r = n_r - k_r $ and $m_c = n_c - k_c $.

When one block is used as one Raptor symbol (i.e. $N_i$=1), at the outer encoder, for every $K$ source symbols ($K n_B$ source bits), $N-K$ Raptor parity symbols are first generated as described in Section \ref{Ss:Raptor enc./dec.}.
The resulting $n_B$-bit symbols are then arranged into a $k_{r}^{B}$-row by $k_{c}^{B}$-column array such that $k_r^B$ ($k_c^B$) is $k_r/n_B$ ($k_c/n_B$, respectively). When $N_i$ is greater than one, the numbers of source symbols and parity symbols of the outer Raptor code increase to $N_i N$ and $N_i(N-k)$, respectively. Also, each block $B_{i,j}$ contains $N_i$ Raptor symbols of $n_B/N_i$ bits each.

For each row of $k_{c}^{B}$ blocks, parity bits for a binary BCH code are computed taking
the corresponding $k_{c}^{B}n_B$ bits as the message bits, and these parity bits form the symbols $R_{i}^{r}$ in
Fig. \ref{fig:BW_TPC}. Binary BCH codes for the columns are also constructed in the same fashion.
The overall code rate is the product of two code rates $\frac{k}{n} \frac{K}{N}$, where $k$ is the message length of the inner BW-PC  which is equal to $k_r^B k_c^B  n_B$ and $n$ is the corresponding codeword length which is given by $k_r^B k_c^B n_B + k_r^B  m_r + k_c^B  m_c$.

For decoding, the component row-column BCH decoders perform iterative hard-decision decoding  first.
Each row component code is decoded via normal hard-decision BCH decoding (e.g., Berlekamp-Massey decoding).
The columns decoders are then executed taking the hard decisions generated by the row decoders, completing one round of hard-decision iteration\footnote{Variations are possible on iteration scheduling as well as on the implementation of hard decision feedback 
between the row-column decoders.}.
Iteration continues until all the errors are corrected, or an additional round does not correct the remaining errors.
In the latter case, decoding failure is declared and erasure tags are applied to all symbols that reside in the intersecting blocks of the failed
row and column codes. Raptor decoding is then executed as explained in \ref{Ss:Raptor enc./dec.}.

\subsection{Code Design}
We assume a page size of 8 KBs and
the target page error rate (PER) is set below $10^{-12}$.
The overall code rate is set to 0.93.
Given the user data of $k=2^{16}$ bits, we set out to choose reasonable
values for the six parameters: $n_B,k_r,k_c,t_r,t_c$ and $N$.

First, recall that for the Raptor code the decoding error probability depends on the number of parity symbols.
Given the fairly high-code-rate requirement of practical NAND flash memories,
to realize the full performance potential of Raptor codes, a large number of Raptor symbols are necessary per Raptor codeword.
We shall only consider Raptor codes with at least 1000 symbols. This sets the limit on the size of the Raptor symbol: namely, $n_B=k/N$ should be less than or equal to 64. After the value for $n_B$ is chosen,
$k_r$ and $k_c$ as well as $t_r$ and $t_c$ are chosen to provide strong enough inner ECC protection
while satisfying the overall code rate requirement of 0.93 with a comfortable margin.
Finally, $K$ is chosen that ensures the final target PER while
meeting the overall code rate $R$ of 0.93.

Of the various codes that have been constructed and examined, we include two representative cases in Table \ref{TAB:BTPC}. Each intersection contains $N_i$ Raptor symbols. Increasing
$N_i$ ensures reliable Raptor decoding performance, but the number of the source symbols ($N_i K$)
should also not exceed 8192, the source size of the employed R10 code.
$C_1$ represents the combination of a weak inner code with a strong outer code while 
$C_2$ exemplifies that of a strong inner code with a weak outer code.

\begin{table}[b]
\caption{}{The parameters of Raptor code/BW-PC combinations: the row and column codes are $(n,k,t)$ BCH codes and $N_{r}$ denotes the number of parity blocks for Raptor codes ($N_{r}=N-K$) and the number of parity symbols of Raptor codes is $N_{i}\times N_{r}$.}\vspace{0.05cm}
\begin{center}
\begin{tabular}{c||c|c|c|c|c}\hline
Overall codes 		& $n_{B}$ & $N_{i}$			&Row code		&Column code	&$N_{r}$ \\ \hline \hline
$C_{1}$ 	& 14 		&1			& (996,966,3) & (996,966,3) & 58\\ \hline
$C_{2}$ 	& 56 		&4			& (2026,1960,6) & (1970,1904,6) & 7\\ \hline

\end{tabular}
\label{TAB:BTPC}
\end{center}
\end{table}

\subsection{Performance Analysis and Evaluation}\label{Ss:Analysis_B_TPC}

The analysis of decoding failure can be carried out  by obtaining the probability distribution of the number of erased symbols in
hard-decision iteration decoding of BW-PCs. We make use of the analysis given in \cite{Cho:12ICC} for
the BW-PC failure rate in arriving at an expression for the final Raptor decoder failure rate.

To briefly summarize the analysis of  \cite{Cho:12ICC}, let $\epsilon_{i,j}$ be the error event corresponding to the failure of $i$ rows and $j$ columns and let $P_{i,j}$ denote the corresponding error event probability. Then, the decoding failure probability $P_{F}$ of the inner BW-PC
can be expressed as follows:

\begin{equation}\label{EQ:Decoding fail_TPC1}
P_{F}  = \sum_{(i,j)}P_{i,j},
\end{equation}
with
\begin{equation}\label{EQ:Pij in general}
\begin{aligned}
P_{i,j}  =& {{k_r^B}\choose{i}}{{k_c^B}\choose{j}}\sum_{n_1 =0}^{n_B}\sum_{n_2 =0}^{n_B}\cdots\sum_{n_{i\times j} =0}^{n_B}
 \sum_{n_{\alpha_1} = 0}^{m_r}\cdots \sum_{n_{\alpha_i} = 0}^{m_r}\\
&\sum_{n_{\beta_1} = 0}^{m_c} \cdots \sum_{n_{\beta_j} = 0}^{m_c} {{n_B}\choose{n_1}}{{n_B}\choose{n_2}}\cdots {{n_B}\choose{n_{i\times j}}} \\
&{{m_r}\choose{n_{\alpha_1}}}\cdots {{m_r}\choose{n_{\alpha_i}}}{{m_c}\choose{n_{\beta_1}}}\cdots {{m_c}\choose{n_{\beta_j}}}\\
		&P_{e}^{n^e}(1-P_e)^{n_B\cdot i\cdot j+ i\cdot m_r+ j\cdot m_c-n^e}\cdot 1_{S}
\end{aligned}
\end{equation}
where $P_{e}$ is the average raw BER; $n_1, n_2, \ldots , n_{i\times j}$ are the numbers of errors in the individual
intersections of the failed row and column codes;
$n_{\alpha_l}$ is the number of errors in the parity block $R_{\alpha_l}^r$ in the failed row codes ($l = 1, 2, \ldots, i$);
$n_{\beta_{p}}$ is the number of errors in the parity block $R_{\beta_p}^c$ in the failed column code ($p = 1, 2, \ldots, j$); $n^e$ is the total number of errors in the $i\times j$ intersections plus $i$ row parity blocks and $j$ column parity blocks, i.e., $n^e = \sum_{k=1}^{i\times j}n_{k}+\sum_{l=1}^{i}n_{\alpha_l}+\sum_{p=1}^{j}n_{\beta_p}$;
$1_{A}$ is the indicator function which is `1' if and only if the event A is true; and
$S$ represents the condition that everyone of the $i \cdot j$ particular selected row and column codes fail to decode.

The probability of the most dominant event  $\epsilon_{1,1}$, for example, is given by
\begin{equation}\label{EQ:Decoding fail_TPC}
\begin{aligned}
P_{1,1}  =& k_r^B \times k_c^B \sum_{n_1^e =0}^{n_B} \sum_{n_2^e = 0}^{m_r} \sum_{n_3^e = 0}^{m_c}{{n_B}\choose{n_1}}{{m_r}\choose{n_2}}{{m_c}\choose{n_3}}\\
		&P_{e}^{n^e}(1-P_e)^{n_B+m_r+m_c-n^e}\cdot 1_{S}.
\end{aligned}
\end{equation}
where $n_1^e, n_2^e$, and $n_3^e$ are the number of errors in message block $B_{i,j}$ intersected by a pair of failed row and column codes, the number of errors in the parity block $R_i^r$ in the failed row code, and the number of errors in the parity block $R_j^c$ in the failed column code, respectively. Also, $n^e=n_1^e +n_2^e +n_3^e$ and $S$ is the condition for the decoding failure event : 1) $t_r<n_1^e +n_2^e$ and 2) $t_c<n_1^e +n_3^e$.

In this paper, we have evaluated $P_{i,j}$ for $(i,j) = (1,1), (1,2), (2,1), (2,2), (3,1), (1,3)$. Adding more terms did not
increase the overall decoding failure rate for the BW-PC. Now, the failure rate of Raptor decoding, denoted by $P_{Raptor}$,
can be obtained as
\begin{equation}\label{EQ:P_DF_BTPC}
\begin{aligned}
P_{Raptor} &=  \sum_{l=1}^{l=N}P(E_1 | N_E=l)P(N_E=l)   \\
&= \sum_{i=1}^{k_r^B} \sum_{j=1}^{k_c^B}{\left[\min\{1,2^{-N_{i}(N-K-i\cdot j)}\}\cdot P_{i,j} \right]}
\end{aligned}
\end{equation}
where $N_E$ denotes the number of erasure blocks, which is equal to
the total number of intersections among the failed columns and rows.

\begin{figure}[t]
 \begin{center}
  \includegraphics[width=0.5\textwidth]{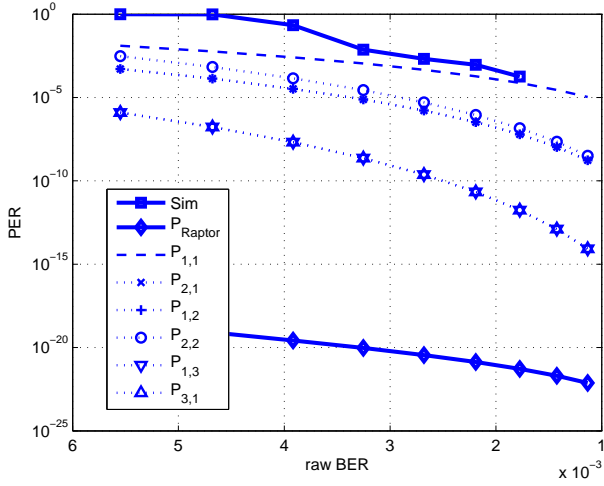}
  \caption{PER simulation and dominant error probability of Raptor codes with BW-PC $C_{1}$}
  \label{FIG:result_C1}
 \end{center}
\end{figure}

\begin{figure}[t]
 \begin{center}
  \includegraphics[width=0.5\textwidth]{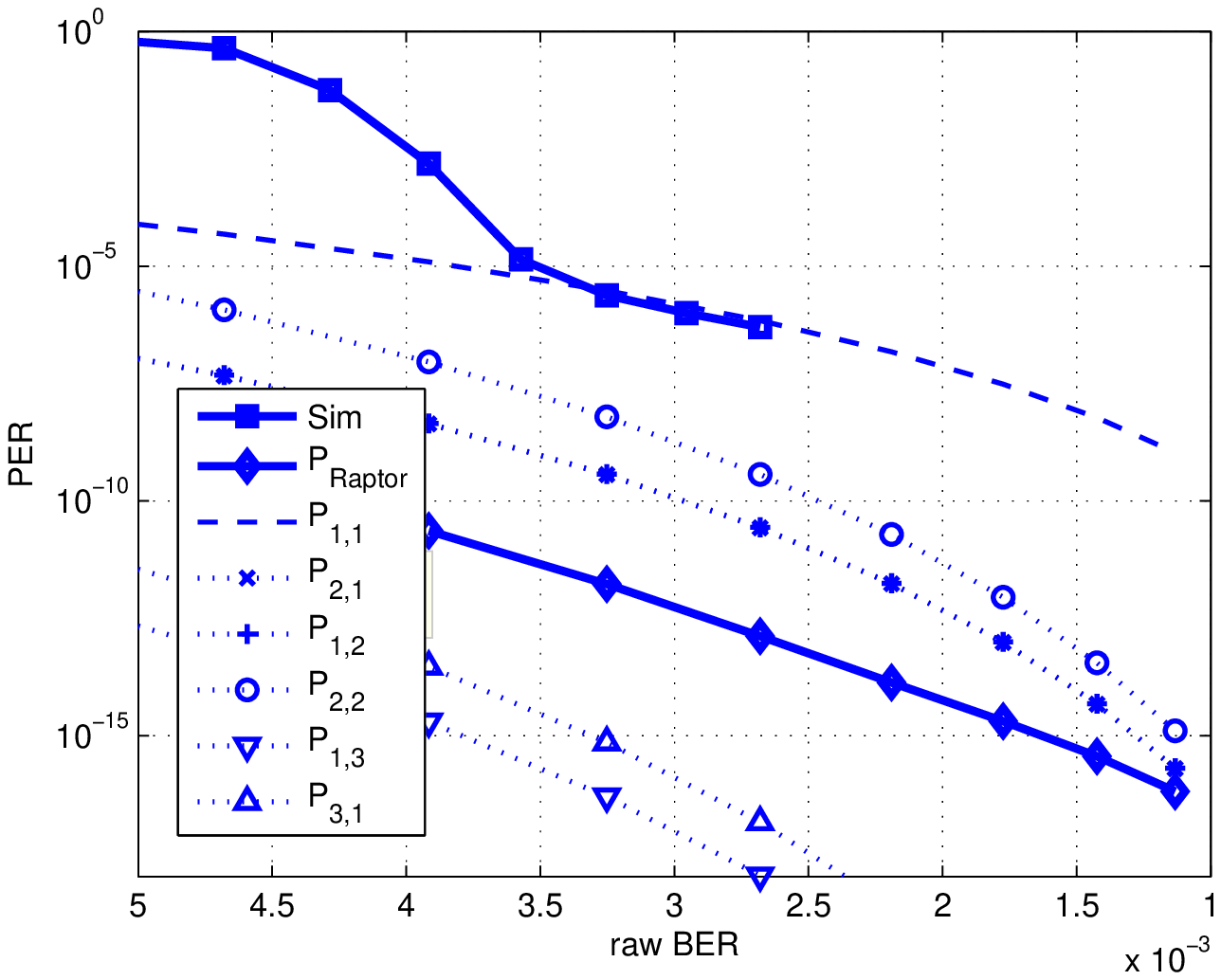}
  \caption{PER simulation and dominant error probability of Raptor codes with BW-PC $C_{2}$}
  \label{FIG:result_C2}
 \end{center}
\end{figure}

The results of the analysis as well as simulations are summarized in Figs. \ref{FIG:result_C1} and \ref{FIG:result_C2}.
In each figure, the curve labeled by `Sim' represents the simulated PERs right after hard-decision iteration decoding is done for
the inner BW-PC. The curves labeled `$P_{i,j}$' are the results of evaluating (\ref{EQ:Pij in general}) for the corresponding error events.
Finally, the curve labeled by `$P_{Raptor}$' is the analysis result corresponding to (\ref{EQ:P_DF_BTPC}).
It can be seen that in each figure, the simulation results for the BW-PC decoding closely follows $P_{1,1}$ in the floor region, indicating that the performance is dominated by the $\epsilon_{1,1}$ event.

Generally speaking, when the inner BW-PC gets strong (as characterized by long codewords and/or low code rates), the probability of dominant error events tend to stay low and the PER curve drops quickly to the error floor. The overall performance indicated by $P_{Raptor}$, however, also depends strongly on the strength of the Raptor code (which improves with increasing number of redundant Raptor symbols $N_iN_s$ and/or decreasing size of the symbol $n_B$).

For example, $C_{1}$ in Fig. \ref{FIG:result_C1} consists of a weak inner BW-PC code and a strong outer Raptor code. In comparison with $C_{2}$ of Fig. \ref{FIG:result_C1}, which combines a strong inner code with a weak outer code,
it can be seen that the $P_{i,j}$ curves for $C_{1}$ are substantially higher,
while the overall performance curve $P_{Raptor}$ appears considerably better with 
$C_{1}$. This is due to the powerful outer code of $C_{1}$ correcting the dominant errors 
that remain after inner decoding. 

Would $C_{1}$ then be a reasonable code, given its superior $P_{Raptor}$ curve that stays well below the target PER at all raw BER values as seen in 
Fig. \ref{FIG:result_C1}? Unfortunately, the answer turns out to be negative, since there were too many miss-correction errors
(i.e., decoding to a valid but incorrect codeword)
observed after inner decoding is done, implying that in this case the analysis predicted by
(\ref{EQ:P_DF_BTPC}) is not reliable. This is due to the fact that in $C_{1}$, the weak component codes
are prone to miss-correction.

\begin{figure}[t]
 \begin{center}
  \includegraphics[width=0.5\textwidth]{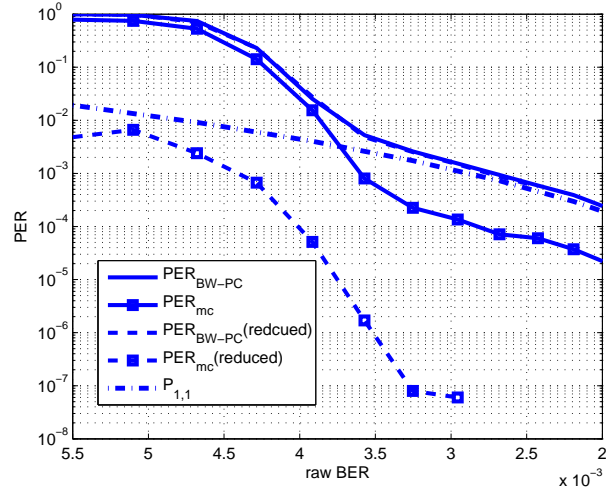}
  \caption{Miss-correction probability (PER$_{mc}$) and decoding failure probability (PER$_\mathbf{BW-PC}$) of an example BW-PC code with weak components codes with and without decoding sphere reduction (from $t$=3 to 2 for each component code). P$_{F}$ is the bound of decoding failure probability of BW-PC obtained in (\ref{EQ:Decoding fail_TPC1}). }
  \label{FIG:Reduced_Comp}
 \end{center}
\end{figure}

The finding that the analytical tool established by (\ref{EQ:P_DF_BTPC}) may be unreliable in certain cases
obviously raises a serious concern regarding the
credibility of the low error rate analysis of (\ref{EQ:P_DF_BTPC}).
In hopes of getting around this issue, for various inner code parameters, 
we checked via simulation for the presence of any miss-correction events
in all row and column codes, using relative weak component codes for which Monte-Carlo 
simulation 
is feasible with reasonable computer time.
Here, we could observe two important trends:

\noindent
1) {After hard-decision-iterative decoding is done, the miss-correction probability curve forms a floor due to the occurrences of some dominant events, as the raw BER is allowed to improve. Specifically, as the channel noise drops and thus the corresponding 
raw BER reduces down to a particular value, dominant miss-correction events 
appear as errors detected by the first component code but not by the second.
Let us call these events \textit{half detected errors}.}

\noindent
2)  {The onset of the floor in the miss-correction rate curve occurs at the same raw BER value as the onset of the floor for the simulated page error rate for the iterative inner decoding. Note that the onset of the floor in the PER curve marks the occurrence of the dominant $\epsilon_{i,j}$ events. In addition, the onset of the floor in the miss-correction rate corresponding to a reduced decoding sphere (the motivation for which is to be given shortly)
 also coincides with that of the PER.} 

Fig. \ref{FIG:Reduced_Comp} shows the simulated decoding failure rates or the PERs after inner iterative decoding as well as the simulated rates of the miss-correction events (denoted by PER$_{mc}$) of
an example BW-PC code with relatively weak components codes given by (1057,1024,3) and (1057,1024,3) with $n_B=16$. We observed that both the decoding failure rate and the miss-correction rate start to get dominated by respective particular events as the raw BER reduces down to 
$3.3\times10^{-3}$. Also seen in Fig. \ref{FIG:Reduced_Comp} is the PER$_{mc}$ corresponding to a case where the decoding sphere is reduced from $t=3$ to $t=2$. The PER 
corresponding to the reduced decoding sphere is increased only slightly (thus nearly indistinguishable in the figure) relative to the PER corresponding to the original decoding sphere set to $t=3$. On the other hand, 
it can be seen that the miss-correction rate for the reduced decoding sphere is improved dramatically.

A reduced decoding sphere provides a means to alleviate the miss-correction issue during inner code decoding. If a \textit{half detected error} event is indicated after the inner BW-PC decoding is completed, a simple strategy is to go back to the inner decoding stage and run the decoders again with a reduced decoding sphere. As the simulation results indicate, this will typically decrease the miss-correction rate by a substantial amount while affecting the decoding failure rate of the overall code very little.   
 
\begin{figure}[t]
 \begin{center}
  \includegraphics[width=0.5\textwidth]{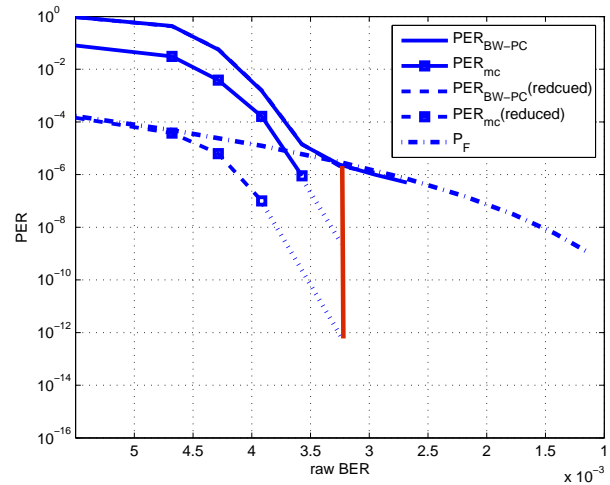}
  \caption{Miss-correction probability (PER$_{mc}$) and decoding failure probability (PER) of the $C_{2}$ code with and without decoding sphere reduction (from $t$=6 to 5 for each component code). The dotted lines are the low-rate extrapolation for PER$_{mc}$ based on the assumption that the onset of the error floor coincide between PER and PER$_{mc}$.}
  \label{FIG:Result_Miss_C2}
 \end{center}
\end{figure}

The trends about the onset of the floor and the dominance of particular events have been observed 
consistently over various BW-PC codes with relatively weak component codes.
If this trend is true for the stronger component codes used in $C_2$, for example, then we can safely conclude that the miss-correction rate of the inner code is suppressed 
to a negligible level using reduced-sphere inner decoding.
In Fig. \ref{FIG:Result_Miss_C2}, simulation results for $C_{2}$'s decoding failure probability (PER) and inner miss-correction probability (PER$_{mc}$) with and without decoding sphere reduction (from $t$ = 6 to 5) are summarized.
 The dotted portions of the PER$_{mc}$ curves represent extrapolation under the assumption that the onset of the floor does not appear until
the PER curve is dominated by the $\epsilon_{i,j}$ events. 
This onset is observed at the raw BER of 3.3$\times 10^{-3}$, as indicated by a vertical line segment. As the page miss-correction rate reaches below $10^{-12}$, the overall target page error rate, we can start to ignore miss-correction events at the inner decoding and the overall code performance will be accurately estimated by (\ref{EQ:P_DF_BTPC}).

We have already seen that $P_{Raptor}$ of $C_{2}$ in Fig. \ref{FIG:result_C1} roughly meets the 
target PER of $10^{-12}$ at the raw BER of 3.3$\times 10^{-3}$. This, combined with the conjecture made in Fig. \ref{FIG:Result_Miss_C2} that the miss-correction rate at inner decoding 
also falls below $10^{-12}$, indicates that the $C_2$ code achieves an acceptable page error rate for raw BERs as high as 3.3$\times 10^{-3}$. 
We again remark that this conjecture 
is based on observing that for relatively weak components codes of various BW-PC inner codes that allow decoding simulation with reasonable computer time,
the onset of the floor for the miss-correction rate curves for the original and reduced decoding spheres coincide with that for the decoding failure rate curve for the hard-decision iterative inner decoding. 

We finally note that when the page size increases and thus even stronger inner component codes can be employed, the problem associated with miss-correction at inner decoding will naturally disappear. The current trend in NAND flash system design is to employ increasingly large page formats which tends to boost system throughput in many applications including mobile devices.
The proposed concatenation scheme would provide an even more attractive ECC option in these applications.

\begin{figure}[t]
 \begin{center}
  \includegraphics[width=0.5\textwidth]{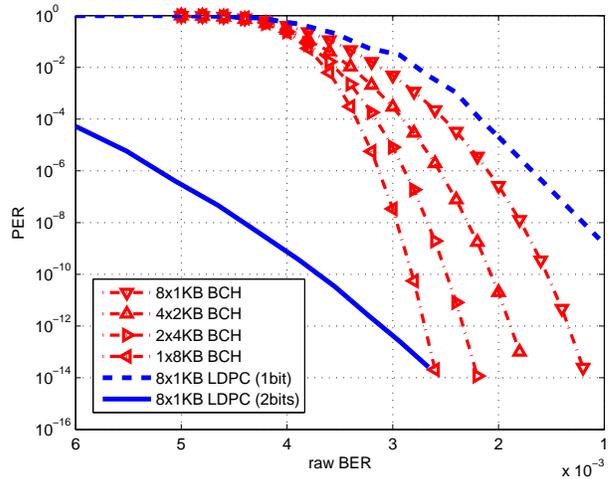}
  \caption{PERs for long BCH codes with different codeword sizes and LDPC codes with limited-precision sensing. PERs of LDPC codes in the high-rate region are evaluated by Monte-Carlo simulation using 1KB LDPC codes and correspond to the probability that any of the 8 LDPC codes in a given page fails. Linear extrapolation (in log-log scale) is used to obtain the PERs below $10^{-6}$.}
 \label{FIG:Result_Comp}
 \end{center}
\end{figure}

\subsection{Performance Comparison with Other Codes}\label{Ss:Performance Comparison}

To compare PER performance of the proposed system with the conventional ECC schemes, we consider the BCH codes of various lengths as well as an LDPC code of length 1 KB.
The results are summarized in Fig. \ref{FIG:Result_Comp}. 
For BCH codes with lengths  1KB - 8 KB, the corresponding word error rates 
have been computed and then weighted by factors 8-1, respectively, to account for the different codeword sizes in evaluating the PER. The codes are specifically (8808, 8192, 44), (17614, 16384, 82), (35232, 32768, 154) and (70534, 65536, 294) BCH codes, all corresponding to a code rate of roughly 0.93. 
Also included are the simulation results for a quasi-cyclic (QC) LDPC code \cite{Fossorier:04JIT,Chen:04JCOM,Fan:00ISTC} with a length of 8249 bits and a variable node degree of 5, a check node degree of 73 and a sub-array size of 113. The progressive-edge-growth (PEG) algorithm of \cite{Hu:GLOBECOM01} is used to generate the particular parity check matrix. 
The LDPC decoder uses the scaled min-sum algorithm \cite{Chen:ISIT05,Zhao:05JCOM} based on 1-bit and 2-bit input quantization. The quantization levels are found based on maximization of mutual information \cite{Jiadong:11GLOBECOM}.
Full internal log-likelihood ratio (LLR) quantization is used for decoding with 
a maximum of 50 iterations. 

By comparing the raw BER values the given codes can tolerate in achieving the $10^{-12}$ PER,
it can be seen that the proposed scheme gives superior performance relative to even the most powerful BCH code of length 8 KB. Compared to the LDPC code (the low rate performance of which has been based on extrapolation assuming no error floor exists), the proposed scheme's performance is comparable to that of LDPC code with 2 bit input-quantization, which would require 3 reads (3 different sensing levels). For higher level quantization, the LDPC code will obviously outperform the propose scheme that relies only on hard-decision sensing. As is widely known, the LDPC code with just 1-bit quantization performs quite poorly. 

From the complexity standpoint, it is well known that the decoding complexity of BCH codes is roughly $O(nt)$ \cite{Hong:Comm95}. Even after considering that the proposed scheme requires 
iteration (2$\sim$3 times on the average), the complexity advantage of our scheme over stand-alone BCH codes becomes obvious mainly due to the much smaller component code sizes. Compared to the LDPC code, again, the performance/complexity tradeoff advantages are clear.

As noted earlier, the RS codes are another clear option for the outer erasure code. 
For example, with a 10-bit symbol [with symbols taken from  $GF(2^{10})$], seven RS codes can be used to 
cover one 8KB page. In this case, the inner BW-PC can be constructed to collect 70 bits or seven RS symbols in each column-row intersection. With proper interleaving, all seven symbols in a given intersection will spread out to seven different RS codewords. Overall, the RS codes in this example can correct eight erasures from the inner BW-PC. While more refined error rate performance comparison will depend on 
detailed inner BW-PC parameter optimization, it is fair to say that this particular RS code configuration
yields results comparable to the Raptor code designs presented in this paper.   

In general, however, the decoding complexity of RS codes is considerably higher than that of Raptor codes. 
It is well known that decoding an RS code in $GF(q)$ requires $O[q (\log{q})^2]$ symbol operations \cite{Justesen:76IT}; in the above scenario RS decoding requires $O(10^5)$ symbol operations (multiplication and addition). 
On the other hand, although hardware implementation options have not been explored extensively in the literature thus far, it is safe to say that Raptor decoding could be done with linear-time complexity of $O(K)$,
as seen in the case of the R10 code \cite{Shokrollahi:11FTCI}. In particular, in the proposed setting 
discussed in this paper, Raptor decoding requires $O(10^3)$ symbol operations (exclusive-ORing). 
It is also worth stressing again that there is a definite trend toward increased page sizes for flash memories 
to improve system throughput \cite{Abraham:12FMS,Vuong:10JEDEC}. With larger page sizes, the performance/complexity tradeoff advantage of Raptor codes over RS codes will become more significant.

\section{Conclusion}\label{S:Conclusion}
This paper has considered application of fixed-rate
Raptor codes to NAND flash memory.
Both an off-line block recovery mode and a real-time page error correction mode have been discussed.
Taking advantage of the linear-time encoding/decoding capability of very long Raptor codes,
significant performance potentials for the BCH-Raptor concatenation have been demonstrated.
The suggested schemes both assume hard-decision sensing of memory cells,
and the inner code's decoder convert the hard-decision input/output storage channel into
a packet-erasure channel for the outer Raptor decoder.
An efficient table-look-up-based encoding/decoding algorithm has been suggested for block recovery that
takes advantage of the unique feature of the storage channel, namely, that the data is stored in the channel medium
for the receiver to access it at any time.
The suggested page-error-correction strategy, on the other hand, utilizes symbol-wise concatenation of
relatively small BCH codes as the inner code, with the intersections of failed row and column codes
tagged as erasure symbols for the outer Raptor code.
Accordingly, the number of shared bits
at the intersection of the row-column codes becomes a key design parameter
for the overall concatenation.
The error rate analysis combined with simulation demonstrate
interesting new possibilities for error control in NAND flash.







\bibliographystyle{IEEEbib}
\bibliography{JSAC_GY_JM}





\end{document}